# Giant tunneling magnetoresistance with a single magnetic phase-transition electrode

Jia Zhang[1*], X. Z. Chen[2], C. Song[2], J. F. Feng[3], H. X. Wei[3], and Jing-Tao Lü[1]

[1]*School of Physics and Wuhan National High Magnetic Field Center, Huazhong University of Science and Technology, 430074 Wuhan, China*

[2]*Key Laboratory of Advanced Materials (MOE), School of Materials Science and Engineering, Tsinghua University, Beijing 100084, China*

[3]*Beijing National Laboratory for Condensed Matter Physics, Institute of Physics, University of Chinese Academy of Sciences, Chinese Academy of Sciences, Beijing 100190, China*

Magnetic phase transition tunneling magnetoresistance (MPT-TMR) effect with a single magnetic electrode has been investigated by first-principles calculations. The calculations show that the MPT-TMR of α'-FeRh/MgO/Cu tunnel junction can be as high as hundreds of percent when the magnetic structure of α'-FeRh changes from G-type antiferromagnetic (GAFM) to ferromagnetic order. This new type of MPT-TMR may be superior to the tunneling anisotropic magnetoresistance because of its huge magneto-resistance effect and similar structural simplicity. The main mechanism for the giant MPT-TMR can be attributed to the formation of interface resonant states at GAFM-FeRh/MgO interface. A direct FeRh/MgO interface is found to be necessary for achieving high MPT-TMR experimentally. Moreover, we find the α'-FeRh/MgO interface with FeRh in ferromagnetic phase has nearly full spin-polarization due to the negligible majority transmission and significantly different Fermi surface of two spin channels. Thus, it may act as a highly efficient and tunable spin-injector. In addition, electric field driven MPT of FeRh-based hetero-magnetic nanostructures can be utilized to design various energy efficient tunnel junction structures and the corresponding lower power consumption devices. We also discuss the consequence of various junction defects on MPT-TMR. The interface oxygen layer is found to be detrimental to MPT-TMR. The sign of MPT-TMR is reversal with Rh termination due to the lack of contribution from interface resonance states. However, the MPT-TMR may be robust against the oxygen vacancy inside of MgO and the shift of Fermi energy. Our results will stimulate further experimental investigations of MPT-TMR and other fascinating phenomenon of FeRh-based tunnel junctions that may be promising in antiferromagnetic spintronics.

# I. INTRODUCTION

There are several types of magnetoresistance effect in magnetic multi-layers including the giant magnetoresistance in ferromagnetic-normal metal-ferromagnetic (FM/NM/FM) metallic multilayer and tunneling magnetoresistance (TMR) in ferromagnetic/insulator/ferromagnetic (FM/I/FM) structures [1-5]. Large magnetoresistance effect at room temperature in these systems leads to rich spintronic device applications, including the read head of HDD, magnetic random access memory (MRAM) and spin-transfer torque MRAM (STT-MRAM) [6]. A typical magnetic tunnel junction (MTJ) consists of two magnetic electrodes separated by an insulating barrier. For MTJ with amorphous $AlO_x$ barrier tens of percent TMR ratio[1,2] and for MTJs with single crystalline barrier hundreds of percent TMR have been achieved experimentally at room temperature[4,5,7,9]. Especially, for MTJ with single-crystalline barrier, the giant TMR is from the combination consequence of spin polarized $\Delta_1$ band of bcc magnetic electrodes and the $\Delta_1$ spin filter effect of the tunnel barrier [1,8,10].

To simplify the MTJ structure, tunneling anisotropic magnetoresistance (TAMR) with only one magnetic electrode has been investigated extensively with various materials, including Fe/GaAs/Au [11-13], CoPt/$AlO_x$/Pt [14]. The TAMR originates from the anisotropic transmission due to the spin-orbit coupling at the magnetic electrode and barrier interface. However, the value of TAMR is rather low and typically less than 1% at room temperature, limiting its practical device application [15].

In this work, we focus on an interesting magnetic material FeRh, which has attracted intensive research interest in spintronics community recently. Here we will focus on the transport properties of a tunnel junction by using FeRh as a magnetic phase transition (MPT) electrode. Bulk cubic FeRh has CsCl-type (B2 or α') structure and undergoes a MPT from G-type anti-ferromagnetic (GAFM) to ferromagnetic (FM) at around 350-370 K [16, 17]. By utilizing its MPT, in the past the α'-FeRh has been considered as materials in thermal assisted magnetic recording [18, 19]. Moreover, recently the FeRh thin film has been shown to have interesting properties, which include the electric-field driven MPT in FeRh/$BaTiO_3$ [20], FeRh/PMN-PT[21] hetero-structures, strain driven MPT in FeRh/MgO[22], electrochemical control of MPT of ultra-thin FeRh thin film[23], room-temperature antiferromagnetic memory resistor[24], voltage controlled magnetic anisotropy[25] *etc*.

Motivated by recent experimental discovery [26], where 20% MPT-TMR at room temperature has been achieved in α'-FeRh/γ-FeRh/MgO/γ-FeRh tunnel junction, we theoretically investigate the transport properties of this new type of TMR effect with single MPT electrode FeRh. As a demonstration and without loss of generality, we investigate the spin-dependent transport properties in α'-FeRh/MgO/Cu tunnel junction by first-principles calculations. Our results show that the MPT-TMR actually can be as high as hundreds of percent. This value is much larger than the experimental and theoretical value of TAMR [11-14], and also much larger than the experimental MPT-TMR value obtained in α'-FeRh/γ-FeRh/MgO/γ-FeRh tunnel junction [26]. This giant MPT-TMR is found to be originating from the interface resonance states formed solely at GAFM-FeRh/MgO interface and thus the direct FeRh/MgO interface is necessary for achieving high MPT-TMR in this type of tunnel junctions. Based on the unique transport properties of α'-FeRh/MgO, we also discuss its possible application as efficient spin-injector and electric-driven tunnel junction beyond the MPT-TMR effect.

## II. CALCULATION METHODS

There are two stable magnetic order in bulk α'-FeRh as illustrated in Fig.1 (a). One is FM phase with magnetic moment on Fe and Rh, the other is GAFM order with equal but opposite magnetic moments on nearest neighbor Fe and zero magnetic moment on Rh. The ground state of bulk-FeRh is GAFM. The energy difference between the FM and GAFM magnetic phase of bulk α'-FeRh has been calculated to be 64.8 meV per formula unit that is close to the theoretical value reported previously [25]. The experimental lattice constant of GAFM-FeRh is 2.99 Å[17], FM-FeRh is 3.01 Å and the lattice constant of MgO is 4.212 Å. The epitaxial relation between FeRh and MgO has been confirmed to be α'-FeRh(001)[110]//MgO(001)[010] by recent experiments [26]. Therefore, the lattice mismatch between MgO and FeRh is small. It is -0.39% and 1.05% for FM-FeRh/MgO and GAFM-FeRh/MgO, respectively. A typical junction structure for FM-FeRh/MgO/Cu is shown in Fig.1(b). The interface structure of GAFM-FeRh/MgO/Cu tunnel junction shown in Fig.1(c), the in-plane unit cell consists of two Fe or Rh atoms per FeRh layer and it is $\sqrt{2} \times \sqrt{2}$ times of FM-FeRh/MgO/Cu junction.

All the calculations in present work are performed by using Quantum-Espresso package [27] with PBE-GGA exchange-correlation potential and ultra-soft pseudo-potential (USPP). The self-consistent calculation of charge density is performed with 15×15×1 $k$-points and plane-wave energy cutoff of 40 Ry. After the electronic structure calculation, the electron transmission calculation is performed by using a standard methodology of wave function scattering method [28]. First, the electronic structure of left electrode FeRh, right electrode Cu and a junction region with FeRh/MgO/Cu supercell are self-consistently calculated. Second, the electron transmission is calculated at different $k_{//}$ points in 2DBZs by matching the wave-function at Fermi energy incident from left across the junction to the right electrode. The transmission is evaluated in 2DBZs with 200×200 $k$-points [28]. The ballistic Landauer conductance of a tunnel junction is proportional to the summation of transmission over $k_{//}$ points in 2DBZ:

$$G^{\uparrow,\downarrow} = \frac{e^2}{h} \sum_{k_{//}} T^{\uparrow,\downarrow}(k_{//})$$

where $T^{\uparrow,\downarrow}(k_{//})$ is the $k$-resolved spin-dependent transmission distribution in ($k_x$, $k_y$) space, $e$ is elementary charge, and $h$ is Plank constant. Similar to the definition of other tunneling magnetoresistance effect, the MPT-TMR can be defined as the relative tunneling conductance ratio when the electrode is at two magnetic phases: MPT-TMR=$G_{GAFM}/G_{FM}-1$, where $G_{GAFM}$ and $G_{FM}$ are the tunneling conductance when the FeRh electrode is in GAFM and FM magnetic phase respectively.

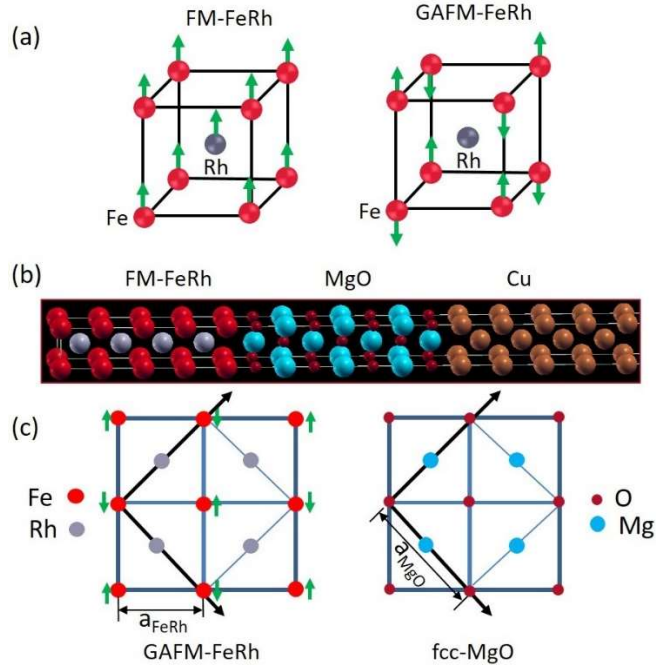

Fig.1 (a) The illustrations of atomic and magnetic structure of bulk FM-FeRh and GAFM-FeRh. (b) The side-viewed structure of FM-FeRh/MgO/Cu tunnel junction (plotted by using XCRYSDEN package [29]). (c) The planar unit cell and epitaxial relation between GAFM-FeRh and fcc-MgO. The green arrows in (a) and (c) indicate the magnetic moment direction.

### III. RESULTS AND DISCUSSIONS

Before discussing the transport properties of the junction, we first look at the spin-dependent Fermi surface (FS) of bulk FeRh shown in Fig. 2 (a) and (b). One could expect a significantly different FS for FM-FeRh and GAFM-FeRh. For the FM-FeRh, the majority and minority channels have quite different FS. Especially, the majority FS has only one available Bloch state, while the minority FS has multiple bands. In addition, there are no available Bloch states around the zone center for majority spin channel and thus one can expect a relative low majority transmission of FM-FeRh through MgO as we will show later. For GAFM-FeRh, the FS of two spins are identical and have multiple Bloch states. The average available Bloch states over 2DBZ of FM-FeRh are larger than that of GAFM-FeRh (see Table 1) and this may lead to a lower resistance in bulk FM-FeRh phase. This is qualitatively consistent with the experimental results [26].

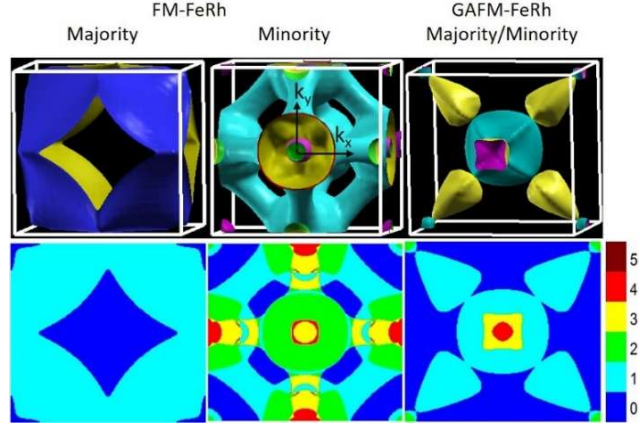

Fig.2 The 3D FS (top panel) and the 2D projected FS (bottom panel) of α'-FeRh in FM and GAFM phase. The color scale in 2D FS indicates the number of available Bloch states at the corresponding $k_{//}=(k_x, k_y)$ points at Fermi level. The 3D Fermi surface of FeRh is plotted by using XCRYSDEN package [29].

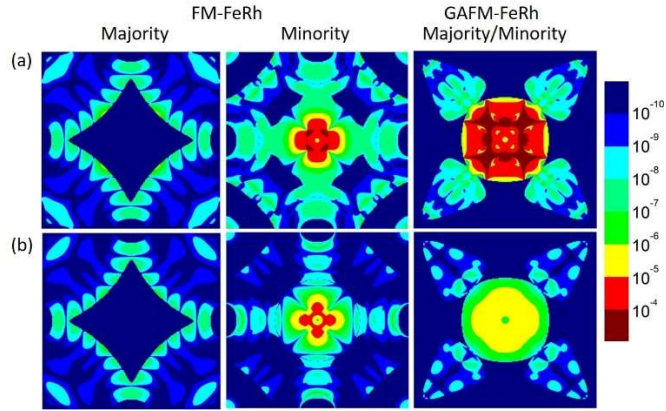

Fig.3 The spin-dependent electron transmission $T(k_{//})$ in 2DBZ with FM and GAFM magnetic order of FeRh for the α'-FeRh/MgO(7 MLs)/Cu junction (a) and α'-FeRh/Cu(1 ML)/MgO(7 MLs)/Cu junction with 1ML of Cu inserted at interface (b).

The electron transmission for α'-FeRh/MgO (7 MLs)/Cu junction is shown in Fig. 3(a). The majority transmission of FM-FeRh/MgO(7 MLs)/Cu junction is relatively low and the largest transmission is at the order of $10^{-8}$, while for minority spin channel the largest transmission is $10^{-5}$~$10^{-4}$ and is distributed around the 2DBZ zone center. For GAFM-FeRh/MgO/Cu tunnel junction, there is a larger area around zone center where transmission is around $10^{-5}$~$10^{-4}$. Consequently, the total transmission of FeRh/MgO/Cu junction with GAFM-FeRh electrode is larger than that with FM-FeRh electrode. The MPT-TMR is calculated to be +562% according to our previous definition. This huge positive MPT-TMR is in contrast to the bulk counterpart where bulk FM-FeRh has higher conductivity than GAFM-FeRh. The large positive MPT-TMR in FeRh/MgO/Cu tunnel junctions implies the fundamental role played by MgO tunnel barrier. In addition, we also calculate the interface ballistic transmission of α'-FeRh/Cu interface without MgO barrier and it turns out that the FM-FeRh/Cu interface has larger transmission than GAFM-FeRh/Cu interface as listed in Table 1. Therefore, there must be some mechanism for the transmission enhancement of GAFM-FeRh/MgO tunnel junction. On the other hand, by recalling the fact that the experimental MPT-TMR in α'-FeRh/γ-FeRh/MgO/γ-FeRh junction with 1ML non-magnetic γ-FeRh at FeRh/MgO interface is only 20% at room temperature which is far below the calculated value of MPT-TMR. These results indicate the crucial role of FeRh/MgO interface for achieving giant MPT-TMR.

Table 1. The total electron transmission over 2DBZ and MPT-TMR of α'-FeRh/MgO(7 MLs)/Cu tunnel junctions with and without 1 ML Cu inserted at FeRh/MgO interface. The experimental tunneling resistance and MTP-TMR in FeRh/γ-FeRh(1u.c.)/MgO(2.7 nm)/γ-FeRh at room temperature, the calculated ballistic transmissions of α'-FeRh/Cu interface and bulk-FeRh and are also listed for reference.

| Structures | FM-FeRh | | GAFM-FeRh | MPT-TMR |
|---|---|---|---|---|
| | Majority | Minority | Majority/Minority[a] | (%) |
| FeRh/MgO(7 MLs)/Cu | $7.89\times10^{-9}$ | $2.28\times10^{-6}$ | $1.51\times10^{-5}$ | +562% |
| FeRh/Cu(1 ML)/MgO(7 MLs)/Cu | $2.95\times10^{-9}$ | $3.98\times10^{-7}$ | $4.61\times10^{-7}$ | +16% |
| FeRh/1u.c.γ-FeRh/MgO(2.7 nm)/γ-FeRh[26] | RA: $3.4\times10^{4}$ (μΩm$^2$) | | RA: $2.8\times10^{4}$ (μΩm$^2$) | +20% |
| FeRh/Cu | 0.70 | 0.12 | 0.24 | -70.4% |
| Bulk-FeRh | 0.78 | 1.46 | 0.60 | -73.2% |

[a]Please note that the majority and minority transmission (conductance) is identical for GAFM-FeRh/MgO/Cu junction and the in plane cell area of GAFM-FeRh is two times of FM-FeRh.

In order to clarify the decisive role of α'-FeRh/MgO interface on MPT-TMR, we look at the transmission with 1ML Cu inserted at the α'-FeRh/MgO interface. The interface metallic layer effect has been investigated in Fe/MgO/Fe-MTJ by previous first-principles calculations [30, 31] and experiments. As shown in Fig.3 (b), after the insertion, the transmission of GAFM-FeRh/MgO/Cu has been drastically suppressed and results in a significant drop of MPT-TMR to 16% (see Table 1 for details). The MPT-TMR now is qualitatively comparable with experimental value in α'-FeRh/γ-FeRh/MgO/γ-FeRh tunnel junction [26]. This proves that the direct interface contact of α'-FeRh with MgO barrier is essential for achieving giant MPT-TMR.

The interface resonance states (IRS) are the states which have bulk origin and the wave function amplitude greatly enhanced at the interface because of interface bonding. The IRS has been already shown to exist in magnetic tunnel junctions for example, in Fe/MgO/Fe-MTJ [32-34], and Fe/GaAs/Cu-MTJ [35]. The presence of IRS has significant contribution to the electron tunneling process in MTJs. In α'-FeRh/MgO/Cu tunnel junction, the giant positive MPT-TMR is also closely related to the IRS formed at the GAFM-FeRh/MgO interface. This can be clearly seen from the layer-resolved density of states (DOS) at FeRh/MgO interface shown in Fig. 4(a). One can immediately see that for GAFM-FeRh/MgO interface there is a pronounced DOS peak appearing at Fermi energy at interfacial Fe of FeRh and O of MgO. In contrast, for FM-FeRh/MgO interface, there is no such feature of IRS. Because of the presence of IRS, the interfacial DOS at the GAFM-FeRh/MgO interface now is much larger than FM-FeRh/MgO interface, which is opposite to the bulk case where the DOS of bulk-FM-FeRh is larger than GAFM-FeRh at Fermi energy. The magnitude of this DOS peak decays away from the interface and reaches a constant value in the bulk region and this is a typical behavior of IRS. One now could naturally understand that by inserting 1ML Cu (or other metallic layers) at interface, the IRS of GAFM-FeRh/MgO will be suppressed and this leads to the reduction of MPT-TMR as we show previously.

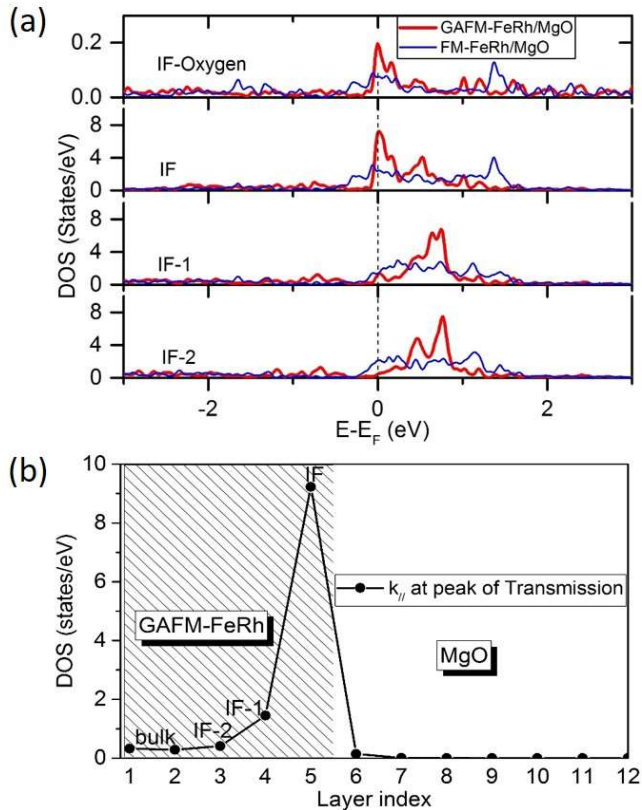

Fig. 4 (a) The minority layer-resolved DOS on Fe layer of FM-FeRh/MgO (blue) and GAFM-FeRh/MgO (red) interfaces. Fermi energy lies at zero as indicated by the dash line. (b) The minority layer resolved DOS at the $k_0$ points $\boldsymbol{k}_0=(k_x, k_y)=(0.19, 0.1)2\pi/a$, where transmission of GAFM-FeRh/MgO is largest. IF, IF-1 and IF-2 means the interface layer, the layers away from the interface by one and two MLs accordingly.

Another direct evidence of the enhanced transmission (conductance) is indeed originating from IRS in GAFM-FeRh/MgO interface is to check if the DOS (wave function amplitude) at the $\boldsymbol{k}_{//}$-points which contribute most of the conductance has been greatly enhanced by the presence of IRS. We plot the layer-resolved DOS at the $k_{//}$ points $k_0=(k_x, k_y)=(0.19, 0.1)$ $2\pi/a$ ($a$ is the in-plane lattice constant) where the transmission is largest in GAFM-FeRh/MgO/Cu tunnel junction. We can see from Fig. 4 (b) the layer-resolved DOS at this $k_{//}$ point greatly increases at the interface and decays rapidly into bulk and finally approaches to a constant value. This is again a typical feature of IRS at this $k_{//}$ point and thus it proves the direct connection of enhanced transmission and IRS in GAFM-FeRh/MgO/Cu tunnel junction.

The MPT-TMR for α'-FeRh/MgO/Cu as a function of MgO thickness is shown in Fig.5. The MPT-TMR can be hundreds of percent which is much larger than TAMR [11-15] and comparable with TMR in Fe/MgO/Fe-MTTs [3-8]. This huge MPT-TMR in α'-FeRh/MgO/Cu make this tunnel junction structure promising for device application at room temperature [26]. Please note that the MPT-TMR for 5ML-MgO is +918% which is slightly smaller than the theoretical value reported +1161% in reference [26]. This is because here we are using a different in-plane cell from the calculations in reference [26] and a fully relaxed interface structure.

The total transmission in 2DBZ as a function of MgO barrier thickness can be well described by an exponential function as it is shown in Fig. 5. For tunnel junctions with thick MgO barrier, the transmission $T$ as a function of MgO barrier thickness $t$ for majority and minority spin channels of FM-FeRh and majority (minority) spin of GAFM-FeRh can be fitted to be $T_{\text{FM-maj}}(t)=10^{0.159-1.18t}$, $T_{\text{FM-min}}(t)=10^{-1.943-0.528t}$, $T_{\text{GAFM-maj/min}}(t)=10^{-1.139-0.526t}$. The majority transmission in FM-FeRh/MgO/Cu junction has relatively larger decay rate while the decay rates of minority transmission in FM-FeRh/MgO/Cu and the transmission in GAFM-FeRh/MgO/Cu tunnel junctions are smaller and close to each other. Therefore, the MPT-TMR ratio will approach to an almost constant value of around 600% with thick MgO limit. This feature is again consistent with an interface dominating effect in MPT-TMR. Meanwhile, it is worthy pointing out that the MgO thickness dependence of MPT-TMR in FeRh-based junction is different from the situation in Fe/MgO/Fe-MTJ where the theoretical ballistic TMR ratio increases with MgO thickness due to the spin-filter effect of MgO barrier [3, 32, 36]. Therefore, in order to achieve sizeable TMR ratio in MgO-based MTJs, the MgO should be thick enough and lead to large $RA$ (resistance-area product) issue for the device application of MTJs [37]. However, this large $RA$ problem may be partly solved in the proposed MPT-TMR structure where thick MgO tunnel barrier is not necessary anymore for obtaining high MPT-TMR value as we discussed previously.

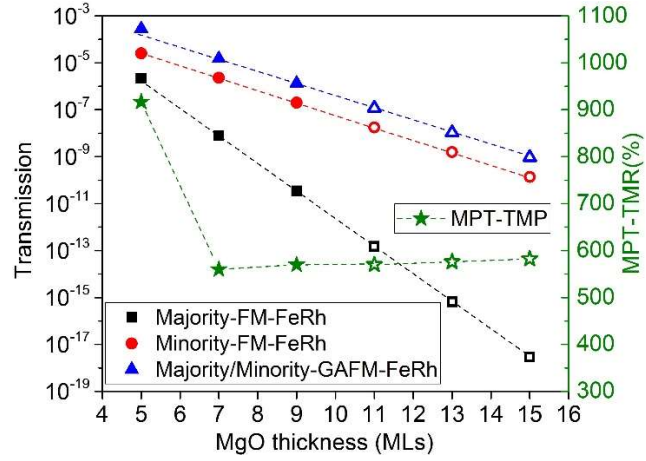

Fig.5 The MgO thickness dependence of transmission (left axis) and MPT-TMR (right axis) for FeRh/MgO/Cu tunnel junctions. The data in solid symbol are the values from first-principles calculation and open symbol are the extrapolated value according to the fitting. The dash lines are guides to eyes.

Our calculations show that giant MPT-TMR can be achieved by MPT of single α'-FeRh magnetic electrode. There are several possible promising applications for this effect. First, as a new TMR effect the giant MPT-TMR effect itself may be attractive for spintronic applications with advantages including simpler structure than typical MTJ and larger TMR than typical TAMR tunnel junctions. Second, besides the giant MPT-TMR, α'-FeRh/MgO can also be used as an efficient spin-injector. One can define the spin-injection polarization as $P=(T_{maj}-T_{min})/(T_{maj}+T_{min})$, where $T_{maj}$ and $T_{min}$ are the transmission through FeRh/MgO interface. In the past, due to its large transmission polarization, (Co)Fe/MgO has been widely used as an efficient spin-injector into semiconductors [38,39]. As we show previously, the transmission of FM-FeRh/MgO is nearly fully spin-polarized and it may be a better spin-injector than (Co) Fe/MgO interface because of the nearly 100% spin-polarization. For example, the spin-injection polarization for FM-FeRh/MgO interface with MgO thickness of 5, 7, 9 MLs are calculated to be 84.5%, 99.3% and 99.9%, respectively. On the other hand, for GAFM-FeRh/MgO, the spin-injection polarization is 0% due to the degeneracy of two spins. Therefore, α'-FeRh/MgO interface can be used as a tunable, high efficient spin-injector and the spin-polarization can be switched between ~100% and 0% upon the magnetic phase transition.

In general, the real interface and junction structure can be complicated due to different defects. In the well-studied Fe/MgO/Fe MTJs, there are various structural defects such as interface roughness, vacancies, contamination and dislocation *etc.*[40]. Those structural defects have important consequence on the spin-dependent transport and the resultant TMR [41-45]. For example, the interface FeO layer was thought to be one of the reason for the lower experimental TMR than theoretical value [41, 42]. However, different from Fe/MgO/Fe MTJs where the lattice mismatch is as high as 4%, the lattice mismatch between α'-FeRh and MgO is quite small and less than 1%. This suggests that the proposed α'-FeRh/MgO/Cu junction may have better epitaxial quality which is confirmed by recent experiments [26]. Nevertheless, below we will briefly discuss the effect of some structural disorder especially the interface imperfection on MPT-TMR.

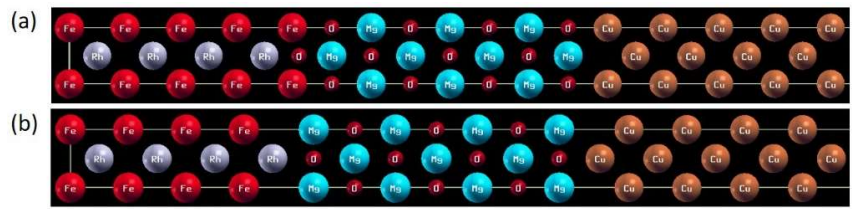

Fig. 6 (a) The α'-FeRh/MgO/Cu tunnel junction with one monolayer of oxygen (FeO) at the interface. (b) The α'-FeRh/MgO/Cu tunnel junction with Rh termination at interface.

First, we consider one monolayer of FeO layer formed at the α'-FeRh/MgO interface. The corresponding junction geometrics are shown in Fig.6 (a). The interfacial oxygen layer is modeled in a similar way to the previous investigation in Fe/MgO/Fe-MTJs [41, 42]. In MgO-MTJs, the interfacial FeO layer is found to be detrimental to the TMR and even the sign of TMR can be reversed [41, 42]. Our calculation results in α'-FeRh/MgO /Cu junction also show similar trend in the proposed MPT-TMR effect. With one monolayer interfacial FeO layer, the MPT-TMR of α'-FeRh/MgO/Cu is down to -3.6% (see transmission listed in Table 2). From the transmission distribution in 2DBZ shown in Fig.7 (a), one can see that with FeO layer at the interface, the transmission for GAFM-FeRh/MgO/Cu has been suppressed and this is responsible for the decrease of MPT-TMR. We also consider transport properties with Rh interface termination as the junction structure is shown in Fig. 6 (b). Despite of the fact that from previous theoretical calculations [25], the ground magnetic phase of α'-FeRh is ferromagnetic with Rh termination and GAFM with Fe termination in FeRh/MgO interface. However, experiments always indicate that the magnetic order in α'-FeRh/MgO interface is GAFM [26]. This suggests the Rh termination interface may not be a favorable interface compared with Fe termination interface. As listed in Table 2, the calculated MPT-TMR with Rh termination is found to be -80% (or equivalently $G_{GAFM}/G_{FM}$=504%) and the sign reversal can be clearly attributed to the lack of IRS contribution to the transmission in GAFM-FeRh/MgO/Cu as shown in Fig.7 (b). At last, we would like to discuss the consequence of interface roughness. The interface roughness should have important influence on IRS and the spin-polarized transport [43, 46]. Here in the case of MPT-TMR, as one may expect the interface roughness may diminish the IRS and therefore it should be detrimental to MPT-TMR and even the sign of MPT-TMR can be reversed. However, a recent work on Fe/MgO/Fe junction shows that the TMR has been surprisingly enhanced with the increasing density of step roughness at Fe/MgO interface [47]. Except the interface disorder, we also consider other possible structural defects for instance the oxygen vacancy inside of MgO barrier, the possible shift of Fermi energy position. In both situations, the MPT-TMR is found to be robust. The detailed discussions can be found in Appendix A. The effect of structural defects in α'-FeRh/MgO-based junction need further experiments and theoretical investigations.

Table 2. The total electron transmission and MPT-TMR of α'-FeRh/MgO(7 MLs)/Cu tunnel junctions with interface FeO layer and Rh termination.

| Structures | FM-FeRh | | GAFM-FeRh | MPT-TMR (%) |
| --- | --- | --- | --- | --- |
| | Majority | Minority | Majority/Minority[a] | |
| FeRh/MgO(7 MLs)/Cu with 1ML Oxygen | $1.21\times10^{-9}$ | $8.34\times10^{-7}$ | $8.04\times10^{-7}$ | -3.6% |
| FeRh /MgO(7 MLs)/Cu with Rh termination | $2.01\times10^{-9}$ | $5.85\times10^{-7}$ | $1.16\times10^{-7}$ | -80% |

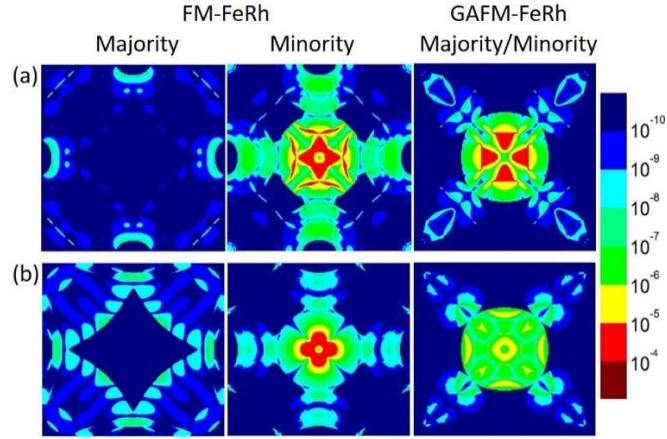

Fig.7 The electron transmission distribution in 2DBZ for (a) α'-FeRh/MgO(7 MLs)/Cu junction with one monolayer of oxygen (FeO) at the interface. (b) α'-FeRh/MgO(7 MLs)/Cu tunnel junction with Rh termination at α'-FeRh/MgO interface.

Applying magnetic field is an ordinary method to control MPT of α'-FeRh electrode and MPT-TMR effect in the proposed tunnel junction. However, magnetic field is a less efficient way to realize MPT-TMR effect since large magnetic field high up to several Tesla is required in order to drive the MPT of α'-FeRh[26]. This issue is also relevant in TAMR structure where large magnetic field is also needed to rotate the direction of magnetic moment [14, 33]. Compared with the conventional MgO-based MTJs, the magnetic field required to switch the magnetic states between parallel and anti-parallel configurations is at the order of hundreds of Oersted at room temperature [4-7]. In this sense, magnetic field driven MPT-TMR may be less practical for real device applications. However, various other energy efficient means including temperature, electric field [20, 21], or ultrafast laser heating can also be adopted to realize MPT-TMR of FeRh-based tunnel junction. For example, the fascinating magnetic phase transition of α'-FeRh in FeRh/BaTiO$_3$ [20] and FeRh/PMN-PT [21] hetero-junctions driven by electric field are recently demonstrated experimentally. By utilizing this MPT of FeRh driven by electric field, an energy efficient MPT-TMR tunnel junction structure can also be designed. For instance, the α'-FeRh/MgO/Cu tunnel junction grown on PMN-PT may have giant MPT-TMR driven by electric field. And also by replacing MgO barrier with ferroelectric insulating materials BaTiO$_3$, in a α'-FeRh/BaTiO$_3$/NM tunnel junction in addition to the electrostatic potential change upon the ferroelectric polarization switch of barrier, there will be an additional MPT in FeRh electrode. This may result in even larger TMR than typical ferroelectric tunnel junctions [47, 48]. Therefore, based on the electric driven MPT of FeRh in the corresponding tunnel junction it may lead to a possible application in low power consumption spintronic devices.

## IV. SUMMARY

In summary, by first-principles calculations, we have investigated the transport properties of a new type of TMR effect with a single MPT electrode α'-FeRh. Our results show that the MPT-TMR effect in α'-FeRh/MgO/Cu tunnel junction can reach hundreds of percent, making it promising as a new TMR effect. The giant MPT-TMR effect arises from the IRS at GAFM-FeRh/MgO interface, which means a clean FeRh/MgO interface is required for achieving high MPT-TMR. We also discuss other possible applications of this interesting tunneling phenomenon especially a highly efficient and tunable spin-injector, and energy efficient electric field driven MPT-TMR in designed tunnel junction structure. Moreover, we discuss the influence of some possible structural defects on MPT-TMR in α'-FeRh/MgO based tunnel junctions. The interface FeO layer is detrimental to MPT-TMR and the tunnel junction with Rh termination interface has negative MPT-TMR due to the lack of IRS contribution. The consequence of atomic roughness should have negative effect on IRS and thus MPT-TMR. And the MPT-TMR is found to be robust against the oxygen vacancy inside of MgO barrier and the position of Fermi energy. However, in real case the resultant transport properties of structural imperfection are needed to be investigated in details. We believe that these results will stimulate further investigation of MPT-TMR effect with α'-FeRh electrode and other possible spintronic device applications of the fascinating phenomenon in various magnetic nano-structures.

## ACKNOWLEDGMENTS

J. Z. is grateful to Prof. X. G. Zhang at University of Florida for fruitful discussions. J. F. F. (Grant No. 51401236), C. S. (Grant No. 51671110), H. X. W. (Grant No. 11674373), J. Z. (Grant No. 11704135) and J. T. L. (Grant No. 61371015) are supported by the National Natural Science Foundation of China. The calculations in this work are performed at National Supercomputer Center in Tianjin, TianHe-1(A) China.

## APPENDIX A: THE EFFECT OF OXYGEN VACANCY AND SHIFT OF FERMI ENERGY ON MPT-TMR

We assume there is oxygen vacancy existing inside of MgO barrier. We model the oxygen vacancy by artificially removing one of oxygen atom in the middle layer of MgO as it is shown in Fig.8 (a). The resultant MPT-TMR has been largely preserved and even enhanced (see transmission listed in Table 3.). This can be seen from Fig.8 (b) where the transmission feature of GAFM-FeRh/MgO/Cu from IRS has been preserved compared with the ideal junction. However, we have to emphasize that besides ballistic transport, there is also diffusive scattering in the junction in the presence of oxygen vacancy, which has not been taken into account in this simple calculation. The diffusive scattering may be important and should be investigated by more advanced calculation method as it is shown in the case of Fe/MgO/Fe-MTJs [45].

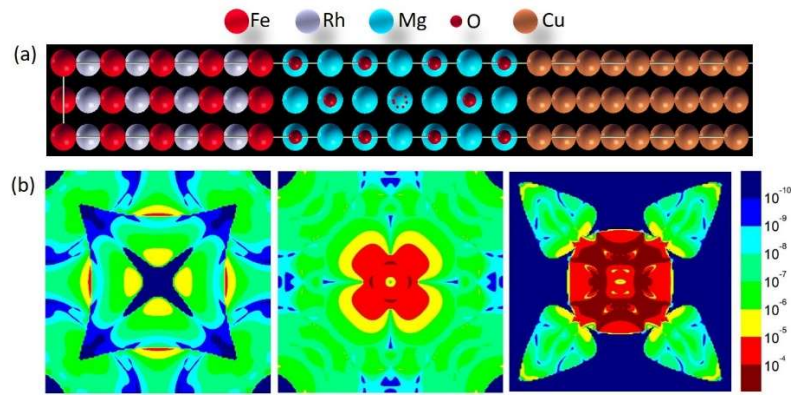

Fig.8 (a) The atomic structure of α'-FeRh/MgO(7 MLs)/Cu junction with one oxygen vacancy in the middle layer of MgO barrier. The dash circle indicates the position of oxygen vacancy. (b) The corresponding electron transmission distribution in 2DBZ.

The position of Fermi energy in real junctions may be different from the ideal junctions because of charge redistribution due to structural disorder and thus it is also important for spin-dependent transport. Especially the IRS contribution may be sensitive to the position of Fermi energy [46]. In order to investigate the effect of the Fermi energy shift, we pick up some energy points respecting to the Fermi energy in the range of $E_F \pm 0.2$ eV. As it is listed in Table 3, our results indicate that in the investigated energy range, the giant MPT-TMR has been largely preserved and hundreds of percent MPT-TMR still can be expected. This feature probably originates from a wide range of resonance peak, as it is evident from Fig.4 (a).

Table 3. The total electron transmission and MPT-TMR of α'-FeRh/MgO(7 MLs)/Cu tunnel junctions with oxygen vacancy and with respecting

to the position of Fermi energy.

| Structures | FM-FeRh | | GAFM-FeRh | MPT-TMR |
|---|---|---|---|---|
| | Majority | Minority | Majority/Minority[a] | (%) |
| FeRh /MgO(7 MLs)/Cu with oxygen vacancy | $3.19 \times 10^{-7}$ | $4.36 \times 10^{-6}$ | $3.48 \times 10^{-5}$ | +1387% |
| $E_F - 0.2$ eV | $2.34 \times 10^{-8}$ | $1.95 \times 10^{-7}$ | $1.97 \times 10^{-6}$ | +802% |
| $E_F - 0.1$ eV | $1.37 \times 10^{-8}$ | $7.27 \times 10^{-7}$ | $4.13 \times 10^{-6}$ | +468% |
| $E_F + 0.1$ eV | $5.08 \times 10^{-9}$ | $4.62 \times 10^{-6}$ | $4.29 \times 10^{-5}$ | +829% |
| $E_F + 0.2$ eV | $3.43 \times 10^{-9}$ | $7.88 \times 10^{-6}$ | $3.58 \times 10^{-5}$ | +354% |

[*]jiazhang@hust.edu.cn;